\documentclass[pdftex,12pt]{article}

\usepackage{amsmath}
\usepackage{amsfonts}
\usepackage{amssymb}
\usepackage{amsthm}
\usepackage{physics}
\usepackage{color}
\usepackage{appendix}
\usepackage{graphicx}
\usepackage{enumitem}
\setenumerate{topsep=3pt,itemsep=2pt}
\setitemize{topsep=3pt,itemsep=2pt}

\usepackage{geometry} % Adjusts page margins
\usepackage{slashed}

\usepackage[colorlinks=true,linkcolor=blue,citecolor=magenta,linktocpage=true]{hyperref}
\usepackage{cite}

\newcommand{\next}[1]{\nonumber \\ &\hspace{#1}}

\usepackage[dvipsnames]{xcolor}
\usepackage{esint}
\usepackage{wrapfig}
\usepackage{bbm}
\newcommand{\one}{\mathbbm{1}}
\newcommand{\mtrx}[1]{{\renewcommand{\arraystretch}{1.0}\begin{pmatrix}#1\end{pmatrix}}}
\usepackage{verbatim}
\newcommand{\pr}[1]{\left(#1\right)}

\newcommand{\pd}{\partial}

\newcommand{\Z}{\mathbb{Z}}
\newcommand{\R}{\mathbb{R}}
\newcommand{\C}{\mathbb{C}}

\newcommand{\PS}{\mathcal{P}}
\newcommand{\HS}{\mathcal{H}}

\newcommand{\Ei}{\mathcal{E}}

\newcommand{\NTr}{\operatorname{NTr}}

\newcommand{\psd}{^{\phantom{.}}}

\usepackage{stmaryrd}
\newcommand{\entry}[3]{\llbracket #1 \rrbracket{}^{#2}{}_{#3}}

\renewcommand{\:}{\hspace{0.10em}}

\begin{document}

\title{A Cubic Matrix Action for the \\ Standard Model and Beyond}
\author{Yigit Yargic\thanks{v-yyargic@microsoft.com} 
$^1$, Jaron Lanier\thanks{jalani@microsoft.com} 
$^1$, Lee Smolin\thanks{lsmolin@perimeterinstitute.ca} 
$^2$, Dave Wecker\thanks{dave.wecker@microsoft.com} 
$^1$}
\date{
$^1$Microsoft Research, Redmond, WA 98052, USA \\[3pt]
$^2$Perimeter Institute for Theoretical Physics, \\
31 Caroline Street North, Waterloo, ON N2J 2Y5, Canada \\[3ex]
January 11, 2022}

\maketitle

\begin{abstract}
    We propose a new framework for matrix theories that are equivalent to field theories on a toroidal spacetime. The correspondence is accomplished via infinite Toeplitz matrices whose entries match the field degrees of freedom on an energy-momentum lattice, thereby replacing the background geometry with matrix indices. These matrix theories can then be embedded into the purely cubic action of a single matrix and combined into a common universality class. We reconstruct the Standard Model action in this framework and discuss its extensions within the same class.
\end{abstract}

\newpage

\tableofcontents

\newpage

\section{Introduction}

Matrix theories have experienced a surge in popularity over the last decades as a promising avenue for fundamental physics \cite{DiFrancesco:1993cyw, Banks:1996vh, Ishibashi:1996xs, Taylor:2001vb, Ambjorn:2008jf, Grosse:2008xr, Steinacker:2010rh, Eichhorn:2013isa, Brahma:2021tkh, Smolin:2000fr, Smolin:2001hh, Alexander:2021rch}, alongside their ubiquity in computational algorithms, numerical simulations, and machine learning. Non-perturbative calculations in quantum field theory (QFT) often rely on the lattice field theory approach, in which the continuum dynamics of a physical system is approximated via a discretization of the underlying spacetime that allows the use of matrix tools \cite{Bongiovanni:2015kia, Banuls:2019rao, deForcrand:2009zkb}.

We present in this paper a novel approach for turning any QFT into a matrix theory. Here, we compactify the spacetime into a torus, thereby discretizing the energy-momentum space rather than the spacetime \cite{Kazakov:2000ar}. The field degrees of freedom are organized into infinite Toeplitz matrices, and the spacetime dependence of fields is exchanged entirely with matrix indices. The spacetime derivatives are expressed via the commutator with a set of constant matrices, and they recreate a precise notion of locality on the spacetime continuum for these pure matrix models.

Our matrix-based approach establishes a new computational framework for field theory. Unlike lattice field theory, it preserves the continuity of the spacetime, and thus the local geometric and gauge symmetries of a system. Consequently, it also evades the fermion-doubling problem in lattice gauge theory \cite{Zielinski:2016qxi}. Hence, this matrix approach to field theory promises some essential advantages and reliability when applying raw computer power to calculations in high-energy physics.

In addition to these computational opportunities, the matrix framework provides an organization scheme for certain sets of QFTs with different degrees of freedom into a structured theory space. From the cubic action $S \sim \Tr(M^3)$ of a single matrix $M$, we can generate different QFTs by a series of choices about including versus excluding parts of the matrix $M$ as degrees of freedom in the theory. We call these choices a \emph{freezing}. Expanding upon the results in \cite{Smolin:2008pk}, we show here through explicit construction that the universality class of theories which can be expressed through a freezing of the cubic matrix action includes the entire Standard Model (SM).

The purely cubic ansatz for the action is associated with background independence \cite{Smolin:2000kc}, and it is motivated in topological field theory \cite{Ishii:2007sy} and string field theory \cite{PhysRevLett.57.283}. While converting any given field theory into a matrix theory in our framework is straightforward, finding a freezing to embed the theory into a cubic structure is often tricky. These challenges can be turned into a new concept of \emph{naturalness} with explicit criteria to assess and motivate field theories within the cubic matrix framework.

Chern-Simons, Abelian BF, and Dirac actions are examples of theories that score high with respect to our cubic naturalness criteria, whereas Yang-Mills theory requires a newfound fine-tuning in this formulation. More generally, the cubic matrix freezing provides a change in perspective to the standard formulation of field theories, which may both introduce new fine-tuning problems that did not exist previously, and offer solutions to the existing ones. We propose here turning these challenges into a tool to motivate, build, and search for new theories in fundamental physics.

For the purposes of this paper, we restrict our attention to classical field theories with a Lagrangian formulation on a toroidal spacetime, and we provide a pure matrix reformulation for such classical actions. We will address the quantization of these matrix actions in a subsequent paper (see also \cite{Adler:2002fu, Minic:1997ek, Starodubtsev:2002fk}). The path integral formalism for matrix theories is an active subject in the literature \cite{Anninos:2020ccj, Lin:2020mme}, however many known results in this context do not apply immediately to the models here due to our distinct approach to the regularization of infinities.

In particular, we do \emph{not} follow here the common approach \cite{HOOFT1974461, PhysRevLett.48.1063, Taylor:1997dy, Steinacker:2003sd} to $U(N)$ matrix models, which is often considered in the limit $N \rightarrow \infty$, but we instead develop a new model for field theories that correspond to the constituents of the Standard Model. In addition, we introduce our matrices as linear operators on an infinite-dimensional Hilbert space together with the Toeplitz property \cite{ToeplitzZurTD} with respect to a fixed basis, instead of relying on an infinite-size limit of finite matrices.

To formulate an action from infinite Toeplitz matrices, we define a linear map called \emph{NTrace} exclusively for Toeplitz matrices, which behaves as a normalized but basis-dependent version of the trace operation. In comparison, the normalization removes the scaling of trace with the matrix size which plays an important role in most other types of matrix models. The basis-dependence implies that the NTrace lacks the cyclicity property of the trace, leading to different combinatorial factors than in the finite trace models. These features are essential to the consistency of the field-theoretical matrix models here, but they require special care when comparing our approach to other matrix theories in the literature (see e.g.~\cite{Marino, Subjakova:2020prh}).

The paper is organized as follows: We establish in Section \ref{sec:Fields} the relationship between the degrees of freedom in a classical field theory and a theory of Toeplitz matrices with the NTrace, providing an equivalence between the two types of theories. In Section \ref{sec:Freezing}, we explain the use of the freezing technique for theory-building on a finite toy model. We apply this technique in Section \ref{sec:Cubic} to our Toeplitz models and embed several field theories into the cubic matrix framework. These results are put together in Section \ref{sec:SM} to express the Standard Model as a freezing of the pure, cubic matrix action. We elaborate in Section \ref{sec:BSM} the concept of naturalness in the freezing approach and its use for discussing theories beyond the Standard Model, before we conclude in Section \ref{sec:Conclusion}.

\section{Matrix formulation of field theories}
\label{sec:Fields}

In this section, we present how the infinitely many degrees of freedom of a classical field can be organized into a pure matrix structure, and how the derivatives and local actions arise in this framework. We consider here classical field theories on a flat, toroidal spacetime in arbitrary signature. Without loss of generality, we take the radii of the spacetime torus to be 1 in a fixed coordinate frame.

\subsection{Field matrices}

Let us consider a scalar field $\phi(x)$ with $x \in [0, 2\pi)^d$ in $d$ dimensions, which has 1 degree of freedom per point of the background geometry. The bridge between this traditional description of a classical field and our matrix framework goes through the Fourier transform, $\phi(x) = \sum\nolimits_{p \in \Z^d} e^{-ipx} \: \phi(p)$, where we have a discrete energy-momentum space $\Z^d$.

Let the entry of a matrix $A$ at the $m$-th row and $n$-th column be denoted as $\entry{A}{m}{n}$. The matrix $A$ is called \emph{Toeplitz} \cite{ToeplitzZurTD,BottcherIntro} if its entries are identical along every diagonal parallel to the main diagonal. In other words, $A$ is Toeplitz if and only if $j - k = j' - k'$ implies $\entry{A}{j}{k} = \entry{A}{j'}{k'}$ for all $j, k, j', k'$ from the index set of $A$.

The infinite Toeplitz matrices are defined in the same way, but their properties depend on the index set. For a field theory in $d$ dimensions, our first goal is to reorganize the individual degrees of freedom of every field $\phi(p)$ as the entries of an infinite Toeplitz matrix over the index set $\Z^d$.

We consider the Hilbert space $\HS = \ell^2(\Z^d)$ of square-summable sequences whose indices are labeled in $\Z^d$. Let $\{e_j \in \HS\}_{j \in \Z^d}$ be a fixed orthonormal basis. An arbitrary operator $A$ on $\HS$ can be written in a matrix representation with the entries $\entry{A}{j}{k} = \langle e_j, A e_k \rangle$. The Toeplitz operators on $\HS$ with respect to this fixed basis $\{e_j\}$ are defined similarly as above, and we will from here on say ``Toeplitz matrix'' to always mean an operator on $\HS$ with the Toeplitz property with respect to $\{e_j\}$.

We define the \textbf{\emph{field matrix}} corresponding to a field $\phi(p)$ as a Toeplitz matrix $\phi$, whose entries are given by
\begin{align}
    \entry{\phi}{j}{k} = \phi(p)
    \quad \text{with} \quad p = j - k \in \Z^d
    \;, \qquad \text{for all $j,k \in \Z^d$}
    \;.
\end{align}
For example, the main diagonal of a field matrix $\phi$ consists of identical values, which is the field value $\phi(\vec{0})$ at zero momentum. The subdiagonals of $\phi$ (there are $d$ of them) consist similarly of identical values, which are the field values $\phi(\ldots, 0, 1, 0, \ldots)$ at a next-to-the-origin point in the energy-momentum lattice, etc. The momentum dependence in the field picture, $\phi(p)$, is thus replaced with the index dependence $\entry{\phi}{j}{k}$ in the matrix picture.

For a more general scalar or vector field with Lorentz and/or gauge indices, we assign similarly a separate complex-valued Toeplitz matrix for each of their scalar component, e.g.~$\entry{A^a_\mu}{j}{k} = A^a_\mu(j-k)$. For a fermion field $\psi$, we similarly have a Grassmann-valued Toeplitz matrix for each of its Dirac- or Weyl-component.

An important consequence of the Toeplitz property (specifically over the index set $\Z^d$) is that any two complex-valued Toeplitz matrices commute with each other, $[A, B] = 0$. Similarly, any two Grassmann-valued Toeplitz matrices anti-commute with each other (as fermions do). The product of two Toeplitz matrices is another Toeplitz matrix, whose entries are given by the convolution of the entries of the multiplied matrices,
\begin{align}
\begin{array}{ccc}
    \text{Field matrices} & \quad \leftrightarrow \quad & \text{Fields} \\[5pt]
    \entry{A B}{i}{k} = \sum\limits_{j \in \Z^d} \entry{A}{i}{j} \, \entry{B}{j}{k}
    \;,
    & &
    (AB)(p) = \sum\limits_{q \in \Z^d} A(p-q) \, B(q)
    \;,
\end{array}
\end{align}
where $p = i-k$ and $q=j-k$. Therefore, we can identify a product of field matrices with the Toeplitz matrix corresponding to the product of those fields. In other words, this correspondence between fields and field matrices provides an isomorphism between their algebras.

Note that every entry of a field matrix is one variable number, not a function of position or momentum. To transfer the spacetime derivatives into this framework and create a notion of locality for the matrix degrees of freedom, we introduce a special set of constant matrices called the \textbf{\emph{derivative matrices}} $P_\mu$, for $\mu \in \{0, ..., d-1\}$. These are unbounded operators on $\HS$, defined in the same preferred basis $\{e_j\}$ by their entries as
\begin{align}
\label{Fi-P}
    \entry{P_\mu}{j}{k} = j_\mu \, \delta^j_k
    \;, \qquad \text{for } \; j,k \in \Z^d
    \;,
\end{align}
where $j_\mu$ is the $\mu$-component of the $d$-tuple $j$. In other words, the matrices $P_\mu$ are infinite diagonal matrices, and their entries are the integers arranged in an ascending order. The utility of these matrices in our framework is based on the following observation:

For any field matrix $\phi$, the commutator $[P_\mu, \phi]$ is another Toeplitz matrix and its entries are given by
\begin{align}
\begin{array}{ccc}
    \text{Field matrices} & \quad \leftrightarrow \quad & \text{Fields} \\[5pt]
    \entry{[P_\mu, \phi]}{j}{k} = \pr{j_\mu - k_\mu} \entry{\phi}{j}{k}
    \;,
    & &
    (i \pd_\mu \phi)(p) = p_\mu \, \phi(p)
    \;,
\end{array}
\end{align}
where the two expressions are related again through $p = j-k$. The commutator $[-i P_\mu, \phi]$ is thus identified as the Toeplitz matrix corresponding to the derivative of the field, $\pd_\mu \phi(x)$. In this way, we can express any field theory Lagrangian as a Toeplitz matrix, which would be composed of sums and products of field matrices and their commutators with $P_\mu$.

Before we move on, let us also discuss the transposition of a field matrix. Recall our association $\entry{\phi}{j}{k} = \phi(p)$, $p = j-k$, between the Toeplitz matrix entries and momenta. The transposition of a Toeplitz matrix corresponds to a PT transformation on the field arguments, $\entry{\phi^T}{j}{k} = \entry{\phi}{k}{j} = \phi(-p)$, $p = j-k$. A field matrix is hermitian if and only if it corresponds to a real-valued field on the spacetime. Transposing the field matrices becomes necessary when we have the (complex or Grassmann) conjugate of a field in the action. For a complex scalar field $\phi$, we write
\begin{align}
    \phi^\dagger = (\phi^*)^T
    \;, \qquad
    \entry{\phi^\dagger}{j}{k} = \phi(-p)^*
    \;, \qquad p = j-k
    \;,
\end{align}
and for a Dirac spinor $\psi_\alpha$, $\alpha \in \{1,2,3,4\}$, we write
\begin{align}
    \bar{\psi}_\alpha = (\psi_\beta{}^*)^T \, \gamma^0{}_{\beta\alpha}
    \;, \qquad
    \entry{\bar{\psi}_\alpha}{j}{k} = \psi_\beta(-p)^* \, \gamma^0{}_{\beta\alpha} = \bar{\psi}_\alpha(-p)
    \;, \qquad p = j-k
    \;.
\end{align}
The role of these matrix operations in the action will become clear in the examples below.

\subsection{NTrace}

Once a field theory Lagrangian is expressed as an infinite Toeplitz matrix $\mathcal{L}$, we would like to define the action as the trace of this matrix. However, infinite Toeplitz matrices are not trace-class, since their main diagonal is an infinite and constant sequence with the value $\mathcal{L}(0) \equiv \entry{\mathcal{L}}{j}{j}, \forall j$ (no sum over $j$ here). A naive attempt at taking the trace of $\mathcal{L}$ would give $\Tr(\mathcal{L}) = |\Z^d| \, \mathcal{L}(0)$, where $|\Z^d| = \infty$ is the cardinality of $\Z^d$.

To solve this problem, we introduce a normalized trace operation $\NTr$ over the Toeplitz operators on $\HS$, called \textbf{\emph{NTrace}}, by simply removing this infinite factor. We define
\begin{align}
    \NTr(\mathcal{L}) = \entry{\mathcal{L}}{0}{0}
    \;, \qquad \text{for a Toeplitz operator $\mathcal{L}$ on $\HS$.}
\end{align}
We are now ready to translate any field theory action (on a toroidal spacetime) into this framework of Toeplitz operators. For example, the Klein-Gordon action for a complex scalar field can be written as
\begin{align}
\label{Fi-S-KG}
    S_{\mathrm{KG}} &= \NTr(\eta^{\mu\nu} \: [-i P_\mu, \phi^\dagger] \: [-i P_\nu, \phi] - m^2 \: \phi^\dagger \: \phi)
    \;,
\end{align}
which is equivalent to
\begin{align*}
    S_{\mathrm{KG}}
    %&= \eta^{\mu\nu} \: \entry{[-i P_\mu, \phi^\dagger] \: [-i P_\nu, \phi]}{0}{0} - m^2 \: \entry{\phi^\dagger \: \phi}{0}{0}
    %\displaybreak[0] \\
    &= \eta^{\mu\nu} \sum_{p \in \Z^d} \entry{[-i P_\mu, \phi^\dagger]}{0}{p} \: \entry{[-i P_\nu, \phi]}{p}{0}\psd
    - m^2 \sum_{p \in \Z^d} \entry{\phi^\dagger}{0}{p} \: \entry{\phi}{p}{0}\psd
    \displaybreak[0] \\
    &= \eta^{\mu\nu} \sum_{p \in \Z^d} \pr{i p_\mu \: \phi(p)^*} \pr{-i p_\nu \: \phi(p)}
    - m^2 \sum_{p \in \Z^d} \phi(p)^* \: \phi(p)
    \displaybreak[0] \\
    %&= \sum_{p \in \Z^d} \pr{p^2 - m^2} \phi(p)^* \: \phi(p)
    %\displaybreak[0] \\
    &= \oint_{[0, 2\pi)^d} \frac{\dd^dx}{(2\pi)^d} \pr{\pd^\mu \phi(x)^* \: \pd_\mu \phi(x) - m^2 \: \phi(x)^* \: \phi(x)}
    \;.
\end{align*}
Similarly, the Dirac action for a gauged spinor field is given by
\begin{align}
\label{Fi-S-Dir}
    S_{\mathrm{Dirac}} &= i \gamma^\mu{}_{\alpha\beta} \NTr\!\big(\bar{\psi}_\alpha^a \pr{[-i P_\mu, \psi_\beta^a] - i g \: \tau^j{}_{ab} \: A_\mu^j \: \psi_\beta^b}\!\big)
    \;,
\end{align}
which is equivalent to
\begin{align*}
    S_{\mathrm{Dirac}} &= \gamma^\mu{}_{\alpha\beta} \sum_{p \in \Z^4} \entry{\bar{\psi}_\alpha^a}{0}{p} \: \entry{[P_\mu, \psi_\beta^a]}{p}{0}\psd
    + g \: \gamma^\mu{}_{\alpha\beta} \: \tau^j{}_{ab} \sum_{p,q \in \Z^4} \entry{\bar{\psi}_\alpha^a}{0}{p} \: \entry{A_\mu^j}{p}{q} \: \entry{\psi_\beta^b}{q}{0}\psd
    \displaybreak[0] \\
    &= \gamma^\mu{}_{\alpha\beta} \sum_{p \in \Z^4} \bar{\psi}_\alpha^a(p) \: p_\mu \: \psi_\beta^a(p)
    + g \: \gamma^\mu{}_{\alpha\beta} \: \tau^j{}_{ab} \sum_{p,q \in \Z^4} \bar{\psi}_\alpha^a(p) \: A_\mu^j(p-q) \: \psi_\beta^b(q)
    \displaybreak[0] \\
    &= \oint_{[0, 2\pi)^4} \frac{\dd^4x}{(2\pi)^4} \pr{
    i \bar{\psi}_\alpha^a(x) \, \gamma^\mu{}_{\alpha\beta} \pr{
    \pd_\mu \psi_\beta^a(x) - i g \: \tau^j{}_{ab} \: A_\mu^j(x) \: \psi_\beta^b(x)
    }}
    \;.
\end{align*}
The Yang-Mills action becomes
\begin{align}
\label{Fi-S-YM}
    S_{\mathrm{YM}} &= - \frac{1}{4} \NTr(F^{j\mu\nu} F^j_{\mu\nu})
    \;, \qquad
    F^j_{\mu\nu} = [-i P_\mu\psd, A^j_\nu] - [-i P_\nu\psd, A^j_\mu] + g f^{jkl} A^k_\mu A^l_\nu
    \;,
\end{align}
where $F^j_{\mu\nu}$ is the Toeplitz matrix for the field strength tensor with
\begin{align*}
    \entry{F^j_{\mu\nu}}{p}{0}
    &= -i p_\mu \: A^j_\nu(p) + i p_\nu \: A^j_\mu(p) + g f^{jkl} \sum_{q \in \Z^4} A^k_\mu(p-q) \: A^l_\nu(q)
    \;,
\end{align*}
and the equivalence of the action follows similarly. Any classical field action can be converted analogously into a matrix action with Toeplitz matrices and the NTrace.

The domain of the NTrace and its algebraic properties plays a crucial role in the consistency of these matrix theories. Our definition of NTrace is so far restricted to the Toeplitz operators on $\HS = \ell^2(\Z^d)$ with respect to a fixed basis. A fixed basis on $\HS$ is thus needed as part of the definition of NTrace, which is unlike the regular trace operation. In particular, the NTrace generally lacks cyclicity,
\begin{align}
    \NTr(AB) \neq \NTr(BA)
    \;,
\end{align}
and therefore $\NTr(UMU^\dagger) \neq \NTr(M)$ for unitary operators $U$.

We generalize the NTrace operation to finite-dimensional extensions $\HS_{(n)} = \C^n \times \HS$ of the Hilbert space $\HS = \ell^2(\Z^d)$ under the following axiomatic properties:
\begin{enumerate}
    \item If $K$ is an $n \times n$-matrix and $A$ is a Toeplitz operator on $\HS$, then
    \begin{align}
    \label{Fi-Rule1}
        \NTr(K \otimes A) = \Tr(K) \: \NTr(A)
        \;.
    \end{align}
    \item If $K$ is a traceless $n \times n$-matrix and $A$ is an arbitrary operator on $\HS$, then
    \begin{align}
    \label{Fi-Rule2}
        \NTr(K \otimes A) = 0
        \;.
    \end{align}
    \item The NTrace is $\C$-linear.
\end{enumerate}
We will use these properties extensively in Section \ref{sec:Cubic} when deconstructing the cubic matrix action $\NTr(M^3)$ by its Lorentz, gauge, and Toeplitz indices to turn it into the actions (\ref{Fi-S-KG}-\ref{Fi-S-YM}) that we discussed here.

As we expand the action $\NTr(M^3)$ with a given operator $M$ on $\HS_{(n)}$ using the three rules above, it is far from guaranteed to be able to convert it into an expression $\NTr(\mathcal{L})$ with a Toeplitz operator $\mathcal{L}$ on $\HS$. In many cases, this is simply not possible and the action $\NTr(M^3)$ is undefined. The field theories that we consider here are in fact based on a very restricted class of operators $M$ where the Lagrangian $\mathcal{L}$ turns out to be Toeplitz, and the classical matrix theory is well-defined.

An interesting question is whether it might be possible to extend the domain of the NTrace map to non-Toeplitz operators on $\HS$ through an analytic continuation. We conjecture that the answer is positive and that the resulting matrix theories lie outside the realm of local field theories. This subject is beyond the scope of this paper, and we will address it in a future paper.

\subsection{No fermion doubling}

The fermion doubling is a well-known issue for field theories on a spacetime lattice, see e.g.~\cite{Zielinski:2016qxi} and the references therein. We will mention here briefly how discretizing the energy-momentum space instead of the spacetime in our matrix framework helps us sidestep this issue.

The Dirac action on $\R^4$,
\begin{align}
\label{Fi-FD-S0}
    S_{\psi} = \int \dd^4x \, \dd^4y \; \bar{\psi}(x) \: K(x-y) \: \psi(y)
    \;, \qquad
    K(x-y) = i \gamma^\mu \: (\pd_\mu \delta^4)(x-y)
    \;,
\end{align}
has the propagator
\begin{align}
    \langle \bar{\psi}(x) \: \psi(y) \rangle = -i \int_{\R^4} \frac{\dd^4p}{(2\pi)^4} \: \frac{e^{-ip(x-y)}}{\gamma^\mu p_\mu}
\end{align}
with a single pole at $p = 0$. If the spacetime is discretized to a lattice with the lattice spacing $a$, and $K_{x,y} = \frac{i}{2a} \, \gamma^\mu \pr{\delta_{y, x+a\hat{\mu}} - \delta_{y, x-a\hat{\mu}}}$ as in lattice field theory, the fermion propagator becomes
\begin{align}
    \langle \bar{\psi}(x) \: \psi(y) \rangle = -i \int_{\mathrm{BZ}} \frac{\dd^4p}{(2\pi)^4} \: \frac{e^{-ip(x-y)}}{\frac{1}{a} \: \gamma^\mu \sin(pa\hat{\mu})}
    \;,
\end{align}
which has poles not only at $p = 0$, but also on the edges of the Brillouin zone $\mathrm{BZ} = [-\pi/a, \pi/a)^4$, leading to a doubling of the fermion modes per spacetime dimension. This is the source of the doubling problem for ``na\"ive lattice fermions'', and there are various approaches in the lattice field theory literature to address it \cite{Zielinski:2016qxi}.

When we instead discretize the energy-momentum space into a lattice $\PS = (\tfrac{1}{R} \: \Z)^4$ in our framework, the action becomes
\begin{align}
    S_{\psi} = \oint_{[0,2\pi R)^4} \dd^4x \, \dd^4y \; \bar{\psi}(x) \: K(x-y) \: \psi(y)
\end{align}
with the same $K(x-y)$ as in \eqref{Fi-FD-S0}. Then, we obtain the propagator
\begin{align}
    \langle \bar{\psi}(x) \: \psi(y) \rangle = - \frac{i}{\pr{2\pi R}^4} \, \sum_{p \in \PS} \frac{e^{-ip(x-y)}}{\gamma^\mu p_\mu}
\end{align}
with only a single pole at $p = 0$. Therefore, our matrix approach evades the fermion doubling problem of lattice field theories.

\section{Freezing in matrix theories}
\label{sec:Freezing}

We now consider a finite matrix action as a toy model for this section, and explain the method of \emph{freezing} to generate new theories in the Lagrangian formulation.

Conventionally, all variables that are part of an expression for the action are considered by default as independent degrees of freedom in the theory. Then, finding a new theory accounts to finding a new action with these variables.

For an alternative approach, let us consider the action
\begin{align}
\label{Fr-S1}
    S = \Tr(V(M))
\end{align}
of a single matrix $M$ with some polynomial $V$. We can keep this action and the polynomial $V$ fixed, and still create inequivalent theories through different choices about what part of $M$ should be considered as variable degrees of freedom in the theory and what part should be fixed to a constant value. We refer to this process as \emph{\textbf{freezing}}, and to the constant parts of $M$ as \emph{frozen}.

The frozen parameters in a matrix theory are non-variable in the sense that the classical equations of motion do not include the Euler-Lagrange equations from a variation of the action with respect to these parameters. The freezing method can thus be put forward as creating distinct theories through different, incomplete schemes of variation over a fixed action. The frozen variables serve as placeholders which enable the organization of the degrees of freedom in a theory into a single matrix $M$. This allows us to group together theories with different degrees of freedom into a common universality class \cite{Smolin:2008pk} if they can be organized into a matrix $M$ with the same internal decomposition based on their symmetries.

As an example, let us consider the cubic action $S = \Tr(M^3)$ of a hermitian $4 \times 4$ matrix $M$, which can be decomposed as
\begin{align}
	M = \one_2 \otimes \one_2 \, \varphi
	+ \one_2 \otimes \sigma^j \: \chi^j
	+ \sigma^a \otimes \one_2 \, \xi_a
	+ \sigma^a \otimes \sigma^j \: A^j_a
	\;,
\end{align}
where $\one_2$ is the $2 \times 2$ identity matrix, $\sigma^\bullet$ are the Pauli matrices, and the real parameters $\varphi$, $\chi^j$, $\xi_a$, and $A^j_a$ constitute the $1 + 3 + 3 + 9 = 16$ possible variables in $M$. If all of these parameters are included as variables in the theory $S = \Tr(M^3)$, the resulting equations of motion are equivalent to $M^2 = 0$, which only has the trivial zero solution. However, if we \emph{freeze} $\varphi = \chi^j = \xi_a = 0$, and keep only the 9 degrees of freedom from $A^j_a$ in $M$ as the variables of a theory, the action
\begin{align}
\label{Fr-CS}
    S = \Tr(M^3) = -4 \: \varepsilon^{abc} \: \varepsilon^{jkl} A^j_a \: A^k_b \: A^l_c
\end{align}
does admit non-trivial solutions. This theory is in fact the truncation of $SU(2)$ Chern-Simons theory to a point\footnote{We can obtain the complete Chern-Simons theory on a 3-torus similarly by considering the matrix $M$ as an operator on $\C^2 \times \C^2 \times \ell^2(\Z^3)$, taking $A^j_a$ as variable Toeplitz matrices, and freezing $\varphi = \chi^j = 0$, $\xi_a = P_a$.}, and it serves as a toy model for the general idea of constructing theories through different freezings of the cubic matrix action.

The advantage of the freezing method lies particularly in providing a systematic approach for organizing certain classes of theories into a finite landscape.
\begin{itemize}
    \item[--] In the conventional approach of effective field theory, which can be called ``bottom-up'', we would first choose the set of all variables, e.g.~$\{A^j_a, ...\}$, then create a landscape of theories from the actions containing all possible terms with these degrees of freedom. The size of this landscape would generally be infinite at the classical level, i.e.~without renormalizability constraints, as it happens with the $f(R)$ theories of gravity.
    
    \item[--] In the alternative freezing approach, which can be called ``top-down'', we first choose a matrix action, e.g.~$\NTr(M^3)$, then build a landscape of theories from the different subsets of matrix entries being accounted as degrees of freedom, versus being frozen. In this framework, the most generic form of the matrix $M$ is decomposed as a sum of tensor products of smaller matrices based on an expected set of symmetries. In this way, one finds a finite number of potential sets of variables inside $M$, thus a finite power set of these variables, and a finite number of theories in the corresponding landscape.
\end{itemize}
When we deal with questions such as ``What theories might be considered as well-motivated extensions beyond the Standard Model?'', having a finite set of options within any framework becomes very useful. Once we reformulate the Standard Model action as a certain freezing of the (cubic) matrix action, it provides us a finite list of potential fields and interaction terms that are \emph{not} in the Standard Model, but could easily be made part of it by \emph{unfreezing} them. Then, one can ask about each of these frozen terms whether they might be yet-to-be-discovered elements of Nature. This is going to be our goal in Sections \ref{sec:SM} and \ref{sec:BSM}.

Finally, we would like to remark that the freezing method creates a theory landscape which can be traversed by a computer program algorithmically. It is feasible to build a loss function in the space of field theories based on any set of empirical constraints that would be entered into the program manually. With such a loss function, the program can search efficiently for new theories in the freezing framework, and identify those that might be more relevant in the real world.

\section{The cubic matrix action}
\label{sec:Cubic}

As mentioned, we would like to rebuild various classical field theories in our ``top-down'' freezing approach. We follow the ansatz in \cite{Smolin:2008pk} of a purely cubic action
\begin{align}
\label{Cu-S0}
    S = \NTr(M^3)
\end{align}
for some matrix $M$ to be determined by the freezing. We will construct all field theories of interest here as the freezings of this simple action. We restrict our attention to theories on a toroidal 4D spacetime with Lorentzian signature.

\subsection{Abelian BF theory -- Locators}
\label{sec:BF}

Though the Abelian BF theory is not a part of the Standard Model, which is our main target in this paper, it will serve as a demonstrative example for computing a freezing for the action \eqref{Cu-S0} before we deal with some unique challenges that appear in the (non-Abelian) Yang-Mills theory. We will thus provide more explicit details in this example than in the following ones.

Let the matrix $M$ be an operator on the Hilbert space $\HS^{(8)} = \C^4 \times \C^2 \times \ell^2(\Z^4)$ together with a decomposition based on the symmetry group $SO(3,1) \times SU(2)$. The maximal expression for $M$ (without any freezing) can be written as
\begin{align}
    M &= \sum_{A,J} \boldsymbol{\gamma}_{(A)}\psd \otimes \boldsymbol{\tau}_{(J)}\psd \otimes \boldsymbol{X}_{(A,J)}\psd
    \\ \nonumber
    &= \one_4 \otimes \pr{\one_2 \otimes X_{(1,1)} + \tau^j \otimes X_{(1,2)}{}^j}
    \\ \nonumber &\hspace{0.5cm}
    + \gamma_5 \otimes \pr{\one_2 \otimes X_{(2,1)} + \tau^j \otimes X_{(2,2)}{}^j}
    \\ \nonumber &\hspace{0.5cm}
    + \gamma^\mu \otimes \pr{\one_2 \otimes X_{(3,1)}{}_\mu\psd + \tau^j \otimes X_{(3,2)}{}_\mu^j}
    \\ \nonumber &\hspace{0.5cm}
    + \gamma^\mu \gamma_5 \otimes \pr{\one_2 \otimes X_{(4,1)}{}_\mu\psd + \tau^j \otimes X_{(4,2)}{}_\mu^j}
    \\ \nonumber &\hspace{0.5cm}
    + \gamma^{\mu\nu} \gamma_L \otimes \pr{\one_2 \otimes X_{(5,1)}{}_{\mu\nu}\psd + \tau^j \otimes X_{(5,2)}{}_{\mu\nu}^j}
    \\ \nonumber &\hspace{0.5cm}
    + \gamma^{\mu\nu} \gamma_R \otimes \pr{\one_2 \otimes X_{(6,1)}{}_{\mu\nu}\psd + \tau^j \otimes X_{(6,2)}{}_{\mu\nu}^j}
    \;,
\end{align}
where $\gamma^{\mu\nu} = \frac{i}{2} [\gamma^\mu, \gamma^\nu]$ and $\tau^j = \frac{1}{2} \sigma^j$. The matrices $X_{(A,J)}$ are generic operators on $\HS = \ell^2(\Z^4)$, and their entries are considered as variables. The constant matrices $\boldsymbol{\gamma}_A\psd = (\one_4, \gamma_5, \gamma^\mu, \gamma^{\mu} \gamma_5, \gamma^{\mu\nu} \gamma_L, \gamma^{\mu\nu} \gamma_R)$ and $\boldsymbol{\tau}_J\psd = (\one_2, \tau^j)$, which we collectively refer to as the \textbf{\emph{locators}}, determine the location of each set of variables $X_{(A,J)}$ inside the matrix $M$.

Before we break the $SU(2)$ group into $U(1)$, there are thus 12 sets of possible variables $X_{(A,J)}$ inside the matrix $M$, which can be frozen or unfrozen in various combinations. These selections create a landscape of $2^{12}$ theories, though most of these theories would be uninteresting for various reasons.

To construct the Abelian BF theory, we freeze all possibly-variable parts of $M$ except for $X_{(3,2)}{}^3_\mu$, $X_{(5,2)}{}^3_{\mu\nu}$ and $X_{(6,2)}{}^3_{\mu\nu}$, which are set to hermitian Toeplitz matrices. The remaining (frozen) parts of $M$ are all set to zero, except for $X_{(3,1)}{}_\mu$ whose values are set equal to the constant derivative matrices $P_\mu$ defined in \eqref{Fi-P}. Hence, the matrix $M$ under the Abelian BF freezing becomes
\begin{align}
\label{Cu-BF-M1}
    M = \gamma^\mu \otimes \pr{\one_2 \otimes P_\mu + \tau \otimes A_\mu} + i \gamma^{\mu\nu} \gamma_5 \otimes \tau \otimes B_{\mu\nu}
    \;,
\end{align}
where the locator $\tau = \frac{1}{2} \sigma^3$ serves as a generator of $U(1)$. The third power of this matrix is given by
\begin{align}
\label{Cu-BF-M3}
    M^3 &= \tfrac{1}{4} \: \gamma^{\mu} \gamma^{\nu} \gamma^{\rho} \otimes \one_2 \otimes \pr{4 P_\mu P_\nu P_\rho + P_\mu A_\nu A_\rho + A_\mu P_\nu A_\rho + A_\mu A_\nu P_\rho}
    \displaybreak[0] \nonumber \\ &\hspace{0.5cm}
    + \gamma^{\mu} \gamma^{\nu} \gamma^{\rho} \otimes \tau \otimes \pr{A_\mu P_\nu P_\rho + P_\mu A_\nu P_\rho + P_\mu P_\nu A_\rho + \tfrac{1}{4} \: A_\mu A_\nu A_\rho}
    \displaybreak[0] \nonumber \\ &\hspace{0.5cm}
    - \tfrac{i}{4} \: \gamma^{\mu\nu} \gamma^{\rho\sigma} \gamma^{\kappa\lambda} \gamma_5 \otimes \tau \otimes B_{\mu\nu} B_{\rho\sigma} B_{\kappa\lambda}
    \displaybreak[0] \nonumber \\ &\hspace{0.5cm}
    - \tfrac{1}{4} \: \gamma^{\mu\nu} \gamma^{\rho\sigma} \gamma^{\kappa} \otimes \one_2 \otimes B_{\mu\nu} B_{\rho\sigma} P_{\kappa}
    - \tfrac{1}{4} \: \gamma^{\kappa} \gamma^{\mu\nu} \gamma^{\rho\sigma} \otimes \one_2 \otimes P_{\kappa} B_{\mu\nu} B_{\rho\sigma}
    \displaybreak[0] \nonumber \\ &\hspace{0.5cm}
    + \tfrac{1}{4} \: \gamma^{\rho\sigma} \gamma^{\kappa} \gamma^{\mu\nu} \otimes \one_2 \otimes B_{\rho\sigma} P_{\kappa} B_{\mu\nu}
    \displaybreak[0] \nonumber \\ &\hspace{0.5cm}
    - \tfrac{1}{4} \pr{
    \gamma^{\mu\nu} \gamma^{\rho\sigma} \gamma^{\kappa}
    - \gamma^{\rho\sigma} \gamma^{\kappa} \gamma^{\mu\nu}
    + \gamma^{\kappa} \gamma^{\mu\nu} \gamma^{\rho\sigma}
    } \otimes \tau \otimes B_{\mu\nu} B_{\rho\sigma} A_{\kappa}
    \displaybreak[0] \nonumber \\ &\hspace{0.5cm}
    + i \gamma^{\mu\nu} \gamma_5 \gamma^\rho \gamma^\sigma \otimes \tau \otimes B_{\mu\nu} P_\rho P_\sigma
    + i \gamma^\rho \gamma^\sigma \gamma^{\mu\nu} \gamma_5 \otimes \tau \otimes P_\rho P_\sigma B_{\mu\nu}
    \displaybreak[0] \nonumber \\ &\hspace{0.5cm}
    + i \gamma^\sigma \gamma^{\mu\nu} \gamma_5 \gamma^\rho \otimes \tau \otimes P_\sigma B_{\mu\nu} P_\rho
    \displaybreak[0] \nonumber \\ &\hspace{0.5cm}
    + \tfrac{i}{4} \pr{
    \gamma^{\mu\nu} \gamma_5 \gamma^\rho \gamma^\sigma
    + \gamma^\sigma \gamma^{\mu\nu} \gamma_5 \gamma^\rho
    + \gamma^\rho \gamma^\sigma \gamma^{\mu\nu} \gamma_5
    } \otimes \tau \otimes B_{\mu\nu} A_\rho A_\sigma
    \displaybreak[0] \nonumber \\ &\hspace{0.5cm}
    + \tfrac{i}{4} \: \gamma^{\mu\nu} \gamma_5 \gamma^\rho \gamma^\sigma \otimes \one_2 \otimes \pr{B_{\mu\nu} P_\rho A_\sigma + B_{\mu\nu} A_\rho P_\sigma}
    \displaybreak[0] \nonumber \\ &\hspace{0.5cm}
    + \tfrac{i}{4} \: \gamma^\rho \gamma^\sigma \gamma^{\mu\nu} \gamma_5 \otimes \one_2 \otimes \pr{P_\rho A_\sigma B_{\mu\nu} + A_\rho P_\sigma B_{\mu\nu}}
    \displaybreak[0] \nonumber \\ &\hspace{0.5cm}
    + \tfrac{i}{4} \: \gamma^\sigma \gamma^{\mu\nu} \gamma_5 \gamma^\rho \otimes \one_2 \otimes \pr{A_\sigma B_{\mu\nu} P_\rho + P_\sigma B_{\mu\nu} A_\rho}
    \;.
\end{align}
When we take the NTrace of this lengthy expression for $M^3$, all terms vanish by the axiom \eqref{Fi-Rule2} except for those in the last three lines of \eqref{Cu-BF-M3}. Using the axiom \eqref{Fi-Rule1} in these last three lines, we get
\begin{align}
    \NTr(M^3) &= \frac{i}{4} \Tr(\gamma^{\mu\nu} \gamma_5 \gamma^\rho \gamma^\sigma) \Tr(\one_2) \NTr\!\big(B_{\mu\nu} P_\rho A_\sigma + B_{\mu\nu} A_\rho P_\sigma
    \nonumber \\ &\hspace{1cm}
    + P_\rho A_\sigma B_{\mu\nu} + A_\rho P_\sigma B_{\mu\nu}
    + A_\sigma B_{\mu\nu} P_\rho + P_\sigma B_{\mu\nu} A_\rho
    \big)
    %\nonumber \\
    %&= \frac{i}{4} \pr{-4 \: \varepsilon^{\mu\nu\rho\sigma}} 2 \NTr\!\big(
    %B_{\mu\nu} P_\rho A_\sigma - B_{\mu\nu} A_\sigma P_\rho
    %\nonumber \\ &\hspace{1cm}
    %+ P_\rho A_\sigma B_{\mu\nu} - A_\sigma P_\rho B_{\mu\nu}
    %+ A_\sigma B_{\mu\nu} P_\rho - P_\rho B_{\mu\nu} A_\sigma
    %\big)
    \displaybreak[0] \nonumber \\[5pt]
    &= -2i \: \varepsilon^{\mu\nu\rho\sigma} \NTr\!\big(
    B_{\mu\nu} \: [P_\rho, A_\sigma]
    + [P_\rho, A_\sigma] \: B_{\mu\nu}
    - [P_\rho, A_\sigma B_{\mu\nu}]
    \big)
    \displaybreak[0] \nonumber \\[5pt]
    &= 4 \: \varepsilon^{\mu\nu\rho\sigma} \NTr\!\big(B_{\mu\nu} \: [-i P_\rho, A_\sigma] \big)
    \\[5pt]
    &= 2 \: \varepsilon^{\mu\nu\rho\sigma} \oint \frac{\dd^4x}{(2\pi)^4} \, B_{\mu\nu}(x) \: F_{\rho\sigma}(x)
    \;,
\end{align}
which is precisely the Abelian BF action.

For the remaining examples in this paper, we skip most of the details that we wrote out explicitly in this subsection, and simply state a frozen matrix $M$ and provide the result from evaluating the cubic action $S = \NTr(M^3)$ analogously.

\subsection{Klein-Gordon theory}

For our next example, let us find a freezing for the cubic action \eqref{Cu-S0} that would generate the Klein-Gordon action for a complex scalar field $\Phi^a$ (with $a \in \{1,2\}$) in the fundamental representation of $SU(2)$. This will become directly relevant for constructing the Higgs sector of the Standard Model.

The cubic action \eqref{Cu-S0} can at most be a degree-3 polynomial in the operators on $\HS$, which include both the field matrices and the derivative matrices. The kinetic term in the Klein-Gordon action \eqref{Fi-S-KG} is however quartic in these operators. Therefore, we build here the Klein-Gordon action in the first-order formalism using an auxiliary vector field $\Pi^a_\mu \sim i D_\mu \Phi^a$.

As we observed in the previous subsection, the field matrices that couple with the locator $\tau^j$ give rise to fields in the adjoint representation of $SU(2)$, while those that couple with $\one_2$ give rise to fields in the scalar representation. In order to construct a field $\Phi^a$ in the fundamental representation, we need to extend these locator matrices by another row and column.

Let $\Ei_{m,n}$ denote a single-entry matrix, which has a 1 in the $m$-th row, $n$-th column, and zeros for all other entries. Then, extending the generators $\tau^j$ to $3 \times 3$ matrices with zeros on the 3rd row and column, we can couple the scalar field $\Phi^a$ in the fundamental representation with the locators $\Ei_{a,3}$, and the conjugate field $\Phi^a{}^\dagger$ in the anti-fundamental representation with the locators $\Ei_{3,a}$. We can illustrate these $3 \times 3$ locators of the $SU(2)$ group as
\begin{align}
\renewcommand{\arraystretch}{1.25}
\pr{\begin{array}{c|c}
    \renewcommand{\arraystretch}{1.0}
    \begin{array}{rl}
        \one_2 \; \text{:} & \text{Scalar rep.} \\
        \tau^j \; \text{:} & \text{Adjoint rep.}
    \end{array} &
    \begin{array}{rl}
        \Ei_{a,3} \; \text{:} & \text{Fund. rep.}
    \end{array} \\[7pt] \hline
    \begin{array}{rl}
        \Ei_{3,a} \; \text{:} & \text{Anti-fund. rep.}
    \end{array} &
    \begin{array}{rl}
        \Ei_{3,3} \; \text{:} & \text{Scalar rep.}
    \end{array}
\end{array}}
\;.
\end{align}
We also need a quadratic term $- \NTr(\Pi^a{}^\mu{}^\dagger \, \Pi^a_\mu)$ in the action to be able to integrate out the auxiliary field $\Pi^a_\mu$. For this reason, the matrix $M$ for this freezing has to contain the identity operator $\one$ on $\HS$. Finally, we need to generate connection terms $\tau^j{}_{ab} \: \Pi^a{}^\mu{}^\dagger A^j_\mu \: \Phi^b$ for the covariant derivative.

There are three potential issues in this setup that need to be avoided: First, we might accidentally create a mass term $\NTr(\one \: A^j{}^\mu A^j_\mu)$ for the gauge boson, which would break the symmetry. Secondly, we might create the term $\NTr(\one \: P^\mu P_\mu)$, which would make the action undefined, since $P^\mu P_\mu$ is not Toeplitz. Thirdly, the cubic action might contain terms of the form $\NTr(\Pi^a{}^\mu{}^\dagger \: P_\mu \: \Phi^a)$ where the derivative matrix $P_\mu$ cannot be absorbed into a commutator with field matrices, therefore making the Lagrangian non-Toeplitz and the action undefined.

To avoid these three problematic scenarios, we extend the locator matrices $\tau^j$ by two rows and columns with zeros, rather than only one. We also replace the gamma matrices $\gamma^\mu$ with the $8 \times 8$ matrices $\Gamma^\mu$ as locators, which are defined as
\begin{align}
    \Gamma^\mu = \mtrx{\gamma^\mu & \\ & \gamma^\mu{}^T}
    \;.
\end{align}
Finally, we can build the first-order Klein-Gordon action as a cubic matrix model \eqref{Cu-S0} with the freezing
\begin{align}
\label{Cu-M-KG}
    M &= \frac{1}{2} \, \Gamma^\mu \otimes \pr{\one_2 - \Ei_{3,3} - \Ei_{4,4}} \otimes P_\mu + \frac{1}{3} \, g \, \Gamma^\mu \otimes \tau^j \otimes A^j_\mu
    \nonumber \\ &\hspace{0.5cm}
    + \frac{1}{2} \, \one_8 \otimes \pr{\Ei_{a,3} \otimes \Phi^a + \Ei_{3,a} \otimes \Phi^a{}^\dagger}
    + \frac{1}{4} \, \Gamma^\mu \otimes \pr{\Ei_{a,3} \otimes \Pi_\mu^a + \Ei_{3,a} \otimes \Pi_\mu^a{}^\dagger}
    \nonumber \\ &\hspace{0.5cm}
    - \frac{2}{3} \, \one_8 \otimes \pr{\Ei_{3,3} - \Ei_{4,4}} \otimes \one
    \;.
\end{align}
It is implied that the locator matrices are extended with zero-valued rows and columns whenever it is necessary for the matrix arithmetics. The cubic action \eqref{Cu-S0} from this freezing becomes
\begin{align}
\label{Cu-S-KG}
    S &= \NTr\!\big(
    i \: \Pi^a{}^\mu{}^\dagger \: (D_\mu \Phi^a)
    - i \: (D_\mu \Phi^a)^\dagger \: \Pi^a{}^\mu
    - \Pi^a{}^\mu{}^\dagger \: \Pi^a_\mu
    - 4 \: \Phi^a{}^\dagger \: \Phi^a
    \big)
    \;,
\end{align}
with the Toeplitz matrix $(D_\mu \Phi^a)$ given by
\begin{align}
    (D_\mu \Phi^a) = [-i P_\mu, \Phi^a] - i g \: \tau^j{}_{ab} \: A^j_\mu \: \Phi^b
    \;.
\end{align}
The action \eqref{Cu-S-KG} is the Klein-Gordon action with mass $m = 2$ in the units where the radii of the spacetime torus are equal to 1 and $\hbar = 1$.

We can illustrate the matrix \eqref{Cu-M-KG} with respect to the $4 \times 4$ gauge locators as
\begin{align}
\renewcommand{\arraystretch}{1.25}
\pr{\begin{array}{c|c|c}
    \one_2 \: P_\mu \;,\; \tau^j \: A^j_\mu
    & \Ei_{a,3} \: \Phi^a \;,\; \Ei_{a,3} \: \Pi_\mu^a &
    \\[2pt] \hline
    \Ei_{3,a} \: \Phi^a{}^\dagger \;,\; \Ei_{3,a} \: \Pi_\mu^a{}^\dagger
    & - P_\mu \;,\; \one &
    \\[2pt] \hline
    & & - P_\mu \;,\; - \one
\end{array}}
\;.
\end{align}

\subsection{Dirac theory}

Just as we extended the gauge locators $\tau^j$ with zeros in the previous subsection to construct a freezing for a scalar field, the Dirac action requires us to also extend the gamma locators $\Gamma^\mu$ to a 9th row and column filled with zeros. Then, we can pair the fermionic field matrices $\psi^a_\alpha$, $\alpha \in \{1,\ldots, 4\}$, with the locators $\Ei_{\alpha, 9}$ to get the gamma matrix factor in the Dirac action.

With the matrix
\begin{align}
\label{Cu-M-Di}
    M &= \frac{1}{2} \, \Gamma^\mu \otimes \one_3 \otimes P_\mu - g \, \Gamma^\mu \otimes \tau^j \otimes A^j_\mu
    \nonumber \\ &\hspace{0.5cm}
    + \pr{\Ei_{\alpha,\,9} + \Ei_{9,\,\alpha+4}} \otimes \Ei_{a,3} \otimes \psi^a_\alpha
    + \pr{\Ei_{9,\,\alpha} + \Ei_{\alpha+4,\, 9}} \otimes \Ei_{3,a} \otimes \bar{\psi}^a_\alpha
    \;,
\end{align}
the cubic action \eqref{Cu-M-KG} becomes the Dirac action for a fermion $\psi$ in the fundamental representation of $SU(2)$,
\begin{align}
    S = i \gamma^\mu{}_{\alpha\beta} \NTr\!\big(
    \bar{\psi}^a_\alpha \pr{
    [-i P_\mu, \psi^a_\beta] - i g \: \tau^j{}_{ab} \: A^j_\mu \: \psi^b_\beta
    }\!\big)
    \;.
\end{align}
The Dirac action is also susceptible to the third potential problem we discussed in the previous subsection, which is again avoided here through the locators $\Gamma^\mu$. The ratio $-1/3$ between the coefficients of $A^j_\mu$ in \eqref{Cu-M-KG} and in \eqref{Cu-M-Di} originates from the (anti-)commutativity of complex- vs.~Grassmann-valued Toeplitz matrices:
\begin{align}\begin{split}
    \NTr(\Pi^a{}^\mu{}^\dagger \: A^j_\mu \: \Phi^b + A^j_\mu \: \Phi^b \: \Pi^a{}^\mu{}^\dagger + \Phi^b \: \Pi^a{}^\mu{}^\dagger \: A^j_\mu)
    &= 3 \NTr(\Pi^a{}^\mu{}^\dagger \: A^j_\mu \: \Phi^b)
    \;, \\
    \NTr(\bar{\psi}^a_\alpha \: A^j_\mu \: \psi^b_\beta + A^j_\mu \: \psi^b_\beta \: \bar{\psi}^a_\alpha + \psi^b_\beta \: \bar{\psi}^a_\alpha \: A^j_\mu)
    &= - \NTr(\bar{\psi}^a_\alpha \: A^j_\mu \: \psi^b_\beta)
    \;.
\end{split}\end{align}
The matrix \eqref{Cu-M-Di} can be illustrated in the spacetime and gauge locators as
\begin{align}
\renewcommand{\arraystretch}{1.25}
\pr{\begin{array}{c|c|c}
    \gamma^\mu \: P_\mu \;,\; \gamma^\mu \: A^j_\mu
    & & \Ei_{\alpha,\,9} \, \psi^a_\alpha
    \\ \hline
    & \gamma^\mu{}^T \: P_\mu \;,\; \gamma^\mu{}^T \: A^j_\mu
    & \Ei_{\alpha+4,\,9} \, \bar{\psi}^a_\alpha
    \\ \hline
    \Ei_{9,\,\alpha} \, \bar{\psi}^a_\alpha &
    \Ei_{9,\,\alpha+4} \, \psi^a_\alpha &
\end{array}}
\;,
\end{align}
\begin{align}
\renewcommand{\arraystretch}{1.25}
\pr{\begin{array}{c|c}
    \one_2 \: P_\mu \;,\; \tau^j \: A^j_\mu
    & \Ei_{a,3} \, \psi^a_\alpha
    \\ \hline
    \Ei_{3,a} \, \bar{\psi}^a_\alpha
    & P_\mu
\end{array}}
\;.
\end{align}

\subsection{Yang-Mills theory -- Fragments}
\label{sec:YMF}

Let us now finally discuss the freezing for the Yang-Mills action. To be specific, we will again consider the gauge group $SU(2)$.

Since the action \eqref{Cu-S0} is cubic in the operators on $\HS$, we aim for finding a freezing for the Yang-Mills action in the first-order formalism,
\begin{align}
\label{Cu-S-YM}
    S = \NTr\bigg(
    \frac{1}{4} \, \Xi^j{}^{\mu\nu} \: \Xi^j_{\mu\nu}
    - \frac{1}{2} \, \Xi^j{}^{\mu\nu} \: F^j_{\mu\nu}
    \bigg)
    \;,
\end{align}
where $\Xi^j_{\mu\nu}$ is an auxiliary two-form field matrix, and
\begin{align}
    F^j_{\mu\nu} = [-i P_\mu\psd, A^j_\nu] - [-i P_\nu\psd, A^j_\mu] + g f^{jkl} A^k_\mu \: A^l_\nu
\end{align}
is a Toeplitz matrix for the field strength tensor components.

There are two challenges for this task: The first is to create the quadratic term $\NTr(\Xi^j{}^{\mu\nu} \: \Xi^j_{\mu\nu})$ without simultaneously creating a mass term $\NTr(A^j{}^\mu \: A^j_\mu)$ which would break the symmetry, or a term $\NTr(P^\mu \: P_\mu)$ which would make the entire action ill-defined. The second challenge is to avoid a cubic self-interaction term for the 2-form field $\Xi^j_{\mu\nu}$.

To elaborate on these two issues, let us consider the matrix
\begin{align}
\label{Cu-M0-YM}
    M &= \frac{1}{2} \, \Gamma^\mu \otimes \one_2 \otimes P_\mu
    + \frac{1}{3} \, g \, \Gamma^\mu \otimes \tau^j \otimes A^j_\mu
    - \frac{3}{8} \, g^{-1} \, \Gamma^{\mu\nu} \otimes \tau^j \otimes \Xi^j_{\mu\nu}
    \;,
\end{align}
where
\begin{align}
    \Gamma^{\mu\nu} = \frac{i}{2} \, [\Gamma^\mu, \Gamma^\nu]
    = \mtrx{\gamma^{\mu\nu} & \\ & - \gamma^{\mu\nu}{}^T}
    \;.
\end{align}
Substituting \eqref{Cu-M0-YM} into \eqref{Cu-S0} gives
\begin{align}
    S = - \frac{1}{2} \NTr(\Xi^j{}^{\mu\nu} \: F^j_{\mu\nu})
    - \frac{27}{32} \, g^{-3} f^{jkl} \NTr(\Xi^j{}^\mu{}_\nu \: \Xi^k{}^\nu{}_\rho \: \Xi^l{}^\rho{}_\mu)
    \;.
\end{align}
As mentioned, we are missing here the term $\NTr(\Xi^j{}^{\mu\nu} \: \Xi^j_{\mu\nu})$, and we also have to deal with a non-perturbative, cubic self-interaction term for the auxiliary 2-form. This latter term had not appeared in the BF model of Section \ref{sec:BF} only because we had considered an Abelian gauge group there.

Should this $\NTr(\Xi \: \Xi \: \Xi)$ term be considered as a natural addition to the Yang-Mills and BF theories? Could it be a part of Nature? We will address these questions in Section \ref{sec:BSM-Urbantke}. At the moment, we treat it as an unwanted term for the Yang-Mills theory.

The missing quadratic term $\NTr(\Xi \: \Xi)$ can be compared to the $\NTr(\Pi^\dagger \: \Pi)$ term in the Klein-Gordon action \eqref{Cu-S-KG}. To obtain the latter, we had added the identity operator $\one$ to the matrix \eqref{Cu-M-KG}. However, we had used there the fact that $\Pi$ is in the fundamental representation, and is therefore coupled with the extended locators $\Ei_{a,3}$, in order to avoid the contact between $\one$ and $A$ that would give rise to a mass term. We cannot use the same trick here because $\Xi$ is in the adjoint representation.

Therefore, let us instead consider the possibility that the matrix $M$ of the Yang-Mills freezing might be block-diagonal,
\begin{align}
\label{Cu-Block1}
    M = \mtrx{M_1 & \\ & M_2}
    \;.
\end{align}
In that case, the cubic action \eqref{Cu-S0} decouples into two separate pieces,
\begin{align}
    S = \NTr(M^3) = \NTr(M_1^3) + \NTr(M_2^3)
    \;.
\end{align}
The action $S$ can then be interpreted as being made up of two actions, $S_1 = \NTr(M_1^3)$ and $S_2 = \NTr(M_2^3)$. We can build these two actions separately as two different freezings and then combine them in the block-diagonal form \eqref{Cu-Block1} into a freezing of a larger matrix to obtain $S = S_1 + S_2$.

If the matrix $M$ of the cubic action \eqref{Cu-S0} has a block-diagonal form as in \eqref{Cu-Block1}, we refer to it as \emph{fragmented}, and we call each block $M_1, M_2, ...$ a \textbf{\emph{fragment}} of $M$. Fragments are a powerful tool for finding a freezing of the cubic matrix action that would create a specific action while avoiding undesired cross terms.

However, an over-reliance on fragments is contrary to the core idea of a freezing that we laid out in Section \ref{sec:Freezing}, which is to constrain the landscape of theories in a top-down approach. With a sufficiently large number of fragments and auxiliary fields, every action can be expressed as a freezing, which nullifies the point of formulating a theory as a ``top-down'' freezing. It is thus always favorable to build a theory in the freezing framework with as few fragments as possible.

Under the ansatz of a purely cubic matrix action \eqref{Cu-S0} (or any other similar ansatz), the number of fragments required to construct a theory becomes a measure for its ``naturalness'' in this framework. The fewer fragments there are in a freezing, the more ``natural'' we can call the theory. We discuss this topic in Section \ref{sec:Naturalness}.

Unlike the Klein-Gordon and Dirac actions, which we had been able to build from a single fragment, the Yang-Mills action requires two fragments for obtaining it from a purely cubic action. We can build the action \eqref{Cu-S-YM} in two pieces as
\begin{subequations}
\begin{align}
    S_1 &= - \frac{1}{2} \NTr(\Xi^j{}^{\mu\nu} \: F^j_{\mu\nu})
    - \frac{27}{32} \, g^{-3} f^{jkl} \NTr(\Xi^j{}^\mu{}_\nu \: \Xi^k{}^\nu{}_\rho \: \Xi^l{}^\rho{}_\mu)
    \;, \\
    S_2 &= \frac{1}{4} \NTr(\Xi^j{}^{\mu\nu} \: \Xi^j_{\mu\nu})
    + \frac{27}{32} \, g^{-3} f^{jkl} \NTr(\Xi^j{}^\mu{}_\nu \: \Xi^k{}^\nu{}_\rho \: \Xi^l{}^\rho{}_\mu)
    \;.
\end{align}
\end{subequations}
Hence, we solve both of the issues above: We can safely add the identity operator $\one$ to $M_2$, which does not contain $P_\mu$ or $A^j_\mu$, and we also cancel out the $\NTr(\Xi \: \Xi \: \Xi)$ terms through a fine-tuning. Hence, we build the matrices
\begin{subequations}
\label{Cu-M-YM}
\begin{align}
    M_1 &= \frac{1}{2} \, \Gamma^\mu \otimes \one_2 \otimes P_\mu
    + \frac{1}{3} \, g \, \Gamma^\mu \otimes \tau^j \otimes A^j_\mu
    - \frac{3}{8} \, g^{-1} \, \Gamma^{\mu\nu} \otimes \tau^j \otimes \Xi^j_{\mu\nu}
    \;, \\
\label{Cu-M2-YM}
    M_2 &= \frac{3}{8} \, g^{-1} \, \Gamma^{\mu\nu} \otimes \tau^j \otimes \Xi^j_{\mu\nu}
    + \frac{2}{27} \, g^2 \, \one_8 \otimes \one_2 \otimes \one
    \;,
\end{align}
\end{subequations}
so that the block-diagonal matrix $M$ with the blocks $M_1$ and $M_2$ gives rise to the Yang-Mills action.

An interesting point about the matrices in \eqref{Cu-M-YM} is that the coupling constant $g$ could be freely absorbed into the variable field matrices $A$ and $\Xi$, therefore the value of $g$ is not determined by the coefficients of $A$ and $\Xi$ in \eqref{Cu-M-YM}. Instead, the value of $g$ is determined solely by the coefficient of the constant identity matrix in \eqref{Cu-M2-YM}. Hence, we observe here the existence of a certain limit $g \rightarrow 0$ after $g A \rightarrow A$ and $g^{-1} \Xi \rightarrow *\Xi$, where the quadratic term $\NTr(\Xi \: \Xi)$ in the Yang-Mills action would vanish, and the theory would converge to the non-Abelian BF theory.

\subsection{Quartic potential}

Before we build the Standard Model, let us also finally mention the freezing for a quartic potential of the Higgs field $\Phi^a$,
\begin{align}
    S = \NTr\!\pr{\mu^2 \: \Phi^a{}^\dagger \: \Phi^a - \nu \pr{\Phi^a{}^\dagger \: \Phi^a}^2}
    \;.
\end{align}
Once again, we turn this potential into a cubic polynomial using an auxiliary field matrix $\xi$, so that it becomes
\begin{align}
\label{Cu-Qua2}
    S = \NTr\!\pr{\mu^2 \: \Phi^a{}^\dagger \: \Phi^a - 2 \: \nu \: \xi \: \Phi^a{}^\dagger \: \Phi^a + \nu \: \xi^2}
    \;.
\end{align}
The field $\xi$ is a pure scalar, therefore it has to be paired with identity matrices. The Higgs field $\Phi^a$ is in the fundamental representation of $SU(2)$, therefore it should be coupled with the locator $\Ei_{a,3}$. Let us introduce the $5 \times 5$ matrices
\begin{align}
    \iota_1 = {\footnotesize\mtrx{1\! \\ &\!1\! \\ &&\!0\! \\ &&&\!0\! \\ &&&&\!0}}
    \;, \qquad
    \iota_2 = {\footnotesize\mtrx{0\! \\ &\!0\! \\ &&\!1\! \\ &&&\!1\! \\ &&&&\!0}}
    \;, \qquad
    \iota_3 = 2^{1/3} {\footnotesize\mtrx{0\! \\ &\!0\! \\ &&\!0\! \\ &&&\!0\! \\ &&&&\!1}}
    \;.
\end{align}
When we substitute the matrix
\begin{align}
    M &= \Ei_{a,3} \otimes \Phi^a + \Ei_{3,a} \otimes \Phi^a{}^\dagger
    + \frac{2\nu}{3} \pr{- \iota_2 + \iota_3} \otimes \xi
    \nonumber \\ &\hspace{0.5cm}
    + \frac{\mu^2}{3} \, \iota_1 \otimes \one
    + \frac{3}{16 \nu} \pr{- \iota_1 + \iota_2 + \iota_3} \otimes \one
    \;,
\end{align}
into the cubic action \eqref{Cu-S0}, it gives the desired potential terms \eqref{Cu-Qua2} up to a constant.

\section{Standard Model as a matrix theory}
\label{sec:SM}

We are now ready to reconstruct the entire Standard Model action on a toroidal spacetime as a freezing of the purely cubic matrix action
\begin{align}
\label{SM-S0}
    S = \NTr(M^3)
    \;.
\end{align}
While this reformulation itself is equivalent to the classical Standard Model action (apart from the spacetime being a Lorentzian 4-torus), it provides a novel perspective on the naturalness of the Standard Model, and motivates adding new terms that can then be tested experimentally as we discuss in the next section.

It would be instructive to compare our approach here with that of Grand Unified Theories (GUT). In a GUT, the Standard Model is embedded into a model with a larger and unified gauge symmetry, such as $SU(5)$ or $SO(10)$, which is then broken to create the Standard Model at low energies. Figuratively, a GUT places the Standard Model inside a considerably larger frame than what its original picture requires, and fills the gap with a myriad of new degrees of freedom.

By contrast, we look for a minimal freezing for the Standard Model that keeps the gauge group as $SU(3)_C \times SU(2)_L \times U(1)_Y$. A freezing would be minimal when the locator matrices have the smallest size compatible with symmetry requirements, and the fragmentation is kept at a minimum. The set of new degrees of freedom or interaction terms that could reasonably be added to the Standard Model without expanding this minimal frame is then severely limited. Hence, it becomes possible to view the Standard Model as a nearly complete puzzle in this cubic matrix framework with only a few possible extensions to consider.

\subsection{Locators and notation}

We will rely on a uniform locator structure across all fragments of our SM matrix reformulation for consistency. Each fragment can be written symbolically in the form
\begin{align}
\label{SM-Loc}
    M_n = \sum \boldsymbol{\gamma} \otimes \boldsymbol{\lambda} \otimes \boldsymbol{T} \otimes \boldsymbol{X}
    \;.
\end{align}
Here, $\boldsymbol{X}$ stands for either a field matrix, i.e.~a Toeplitz operator on $\ell^2(\Z^4)$, or the constant derivative matrices $P_\mu$ defined in \eqref{Fi-P}. The finite matrices $\boldsymbol{\gamma}$, $\boldsymbol{\lambda}$ and $\boldsymbol{T}$ stand for the locators assigned to $\boldsymbol{X}$, which determine its Lorentz, chromodynamic and electroweak transformation properties, respectively.

The Lorentz locators $\boldsymbol{\gamma}$ of size $11 = 4 + 4 + 3$ are structured as
\begin{align}
\boldsymbol{\gamma} :
\renewcommand{\arraystretch}{1.25}
\pr{\begin{array}{c|c}
    \begin{array}{rl}
        \one_8 \; \text{:} & \text{Scalars} \\[-2pt]
        \Gamma^\mu \; \text{:} & \text{Vectors} \\[-2pt]
        \Gamma^{\mu\nu} \; \text{:} & \text{2-forms}
    \end{array} &
    \begin{array}{rl}
        \Ei_{\alpha,\,8+g} \; \text{:} & \text{Fermions} \\[-2pt]
        \Ei_{\alpha+4,\,8+g} \; \text{:} & \text{Anti-fermions}
    \end{array} \\[7pt] \hline
    \begin{array}{rl}
        \Ei_{\alpha,\,8+g} \; \text{:} & \text{Anti-fermions} \\[-2pt]
        \Ei_{\alpha+4,\,8+g} \; \text{:} & \text{Fermions}
    \end{array} &
    \begin{array}{rl}
        Y \; \text{:} & \text{Scalars}
    \end{array}
\end{array}}
,
\end{align}
where the index $g \in \{1,2,3\}$ stands for the three fermion generations, and $Y$ are the Yukawa matrices. $\mu \in \{0, \ldots, 3\}$ are the Lorentz indices and $\alpha \in \{1, \ldots, 4\}$ are the Dirac spinor indices. The chiral projectors are defined as
\begin{align}
    \Gamma_5 = \mtrx{\gamma_5 & \\ & \gamma_5}
    \;, \qquad
    \Gamma_L = \frac{1}{2} \pr{\one_8 - \Gamma_5}
    \;, \qquad
    \Gamma_R = \frac{1}{2} \pr{\one_8 + \Gamma_5}
    \;.
\end{align}
Next, the chromodynamic locators $\boldsymbol{\lambda}$ of size $4 = 3 + 1$ are structured as
\begin{align}
\boldsymbol{\lambda} :
\renewcommand{\arraystretch}{1.25}
\pr{\begin{array}{c|c}
    \begin{array}{rl}
        \one_3 \; \text{:} & \text{Scalar rep.} \\[-2pt]
        \lambda^u \; \text{:} & \text{Adjoint rep.}
    \end{array} &
    \begin{array}{rl}
        \Ei_{r,\,4} \; \text{:} & \text{Fundamental rep.} \\[-5pt]
        & \text{(quarks)}
    \end{array} \\[10pt] \hline
    \begin{array}{rl}
        \Ei_{4,\,r} \; \text{:} & \text{Anti-fund. rep.} \\[-5pt]
        & \text{(anti-quarks)}
    \end{array} &
    \begin{array}{rl}
        \Ei_{4,\,4} \; \text{:} & \text{Scalar rep.}
    \end{array}
\end{array}}
,
\end{align}
where $\lambda^u$ are the Gell-Mann matrices\footnote{The Gell-Mann matrices themselves are the generators of the fundamental representation of $SU(3)$, but the field matrices (gluons) assigned to them will transform in the adjoint representation.}, $u \in \{1, \ldots, 8\}$ are the $SU(3)_C$ adjoint indices, and $r \in \{1,2,3\}$ are the color indices.

Finally, the electroweak locators $\boldsymbol{T}$ are of size $5 = 2 + 2 + 1$, and include the matrices
\begin{align}
\label{SM-T6}
    T^j = \pr{\begin{array}{c|c|c}
        \tau^j && \\\hline & 0 & \\\hline && 0
    \end{array}}
    \;, \qquad
    T^6 = \pr{\begin{array}{c|c|c}
        0 && \\\hline & \tau^3 & \\\hline && 0
    \end{array}}
    \;,
\end{align}
as well as $\Ei_{a,\,5}$, $\Ei_{a+2,\,5}$ etc., where $j \in \{1,2,3\}$ are the $SU(2)_L$ adjoint indices, and $a \in \{1,2\}$ are the $SU(2)_L$ fundamental representation indices.

The two fermion species, quarks and leptons, are represented by the Grassmann-valued Toeplitz matrices
\begin{align}
    Q^{ra}_{\alpha g}
    \quad \text{and} \quad
    L^{a}_{\alpha g}
    \;.
\end{align}
Their covariant derivatives are given by the Toeplitz matrices
\begin{subequations}
\label{SM-CovFer}
\begin{align}
    (D_\mu\psd Q^{ra}_{\alpha g}) &= [-i P_\mu\psd, Q^{ra}_{\alpha g}]
    - i g_w \: \gamma_L{}_{\alpha\beta}\psd \: \tau^j{}_{ab} \: W^j_\mu \: Q^{rb}_{\beta g}
    - i g_s \: \lambda^u{}_{rs} \: G^u_\mu \: Q^{sa}_{\alpha g}
    \next{2.6cm}
    - i g' \: \gamma_R{}_{\alpha\beta}\psd \: \tau^3{}_{ab} \: B_\mu\psd \: Q^{rb}_{\beta g}
    - i \: \frac{1}{6} \: g' B_\mu\psd \: Q^{ra}_{\alpha g}
    \;, \\[10pt]
    (D_\mu\psd L^{a}_{\alpha g}) &= [-i P_\mu\psd, L^{a}_{\alpha g}]
    - i g_w \: \gamma_L{}_{\alpha\beta}\psd \: \tau^j{}_{ab} \: W^j_\mu \: L^{b}_{\beta g}
    \next{2.6cm}
    - i g' \: \gamma_R{}_{\alpha\beta}\psd \: \tau^3{}_{ab} \: B_\mu\psd \: L^{b}_{\beta g}
    + i \: \frac{1}{2} \: g' B_\mu\psd \: L^{a}_{\alpha g}
    \;.
\end{align}
\end{subequations}
Similarly, the covariant derivative of the Higgs field is written as
\begin{align}
    (D_\mu \Phi^a) = [-i P_\mu, \Phi^a] - i g_w \: \tau^j{}_{ab} \: W^j_\mu \: \Phi^b - i \: \frac{1}{2} \: g' B_\mu\psd \: \Phi^a
    \;.
\end{align}
We will use the field matrices $\tilde{\Phi}^a = i \sigma^2{}_{ab} \: \Phi^b{}^\dagger$ in the Yukawa sector. For each of the four $3 \times 3$ Yukawa matrices $Y \in \{Y^{(u)}, Y^{(d)}, Y^{(\nu)}, Y^{(e)}\}$, we write $\bar{Y}$ for the block-diagonal $11 \times 11$ matrix
\begin{align}
    \bar{Y} = \pr{\begin{array}{c|c}
        0 & \\\hline & Y
    \end{array}}
    \;.
\end{align}

Finally, the auxiliary 2-forms corresponding to the gauge bosons $G_\mu^u$, $W_\mu^j$, $B_\mu\psd$ are denoted as $\Xi_G{}_{\mu\nu}^u$, $\Xi_W{}_{\mu\nu}^j$, $\Xi_B{}_{\mu\nu}\psd$, respectively.

\subsection{The SM matrix}

We build the Standard Model from a freezing of the cubic action \eqref{SM-S0} given by the matrix $M$ which has five fragments,
\begin{align}
    M = \mtrx{M_1 \\ & M_2 \\ && M_3 \\ &&& M_4 \\ &&&& M_5}
    \;.
\end{align}
These fragments can be written as follows:
\begin{align}
\label{SM-M1}
    M_1 &= \frac{1}{2} \, \Gamma^\mu \otimes \one_4 \otimes \one_5 \otimes P_\mu\psd
    \nonumber \\ &\hspace{0.5cm}
    + \frac{1}{3} \, g_s \, \Gamma^\mu \otimes \lambda^u \otimes \Ei_{5,5} \otimes G^u_\mu
    - \frac{3}{8} \, g_s^{-1} \, \Gamma^{\mu\nu} \otimes \lambda^u \otimes \Ei_{5,5} \otimes \Xi_G{}^u_{\mu\nu}
    \nonumber \\ &\hspace{0.5cm}
    + \frac{1}{3} \, g_w \, \Gamma^\mu \otimes \Ei_{4,4} \otimes T^j \otimes W^j_\mu
    - \frac{3}{8} \, g_w^{-1} \, \Gamma^{\mu\nu} \otimes \Ei_{4,4} \otimes T^j \otimes \Xi_W{}^j_{\mu\nu}
    \nonumber \\ &\hspace{0.5cm}
    + \frac{1}{3} \, g' \, \Gamma^\mu \otimes \Ei_{4,4} \otimes T^6 \otimes B_\mu\psd
    - \frac{3}{8} \, g'{}^{-1} \, \Gamma^{\mu\nu} \otimes \Ei_{4,4} \otimes T^6 \otimes \Xi_B{}_{\mu\nu}\psd
\displaybreak[0] \\[10pt]
\label{SM-M2}
    M_2 &= \frac{3}{8} \, \Gamma^{\mu\nu} \otimes \pr{
    g_s^{-1} \: \lambda^u \otimes \Ei_{5,5} \otimes \Xi_G{}^u_{\mu\nu}
    + \Ei_{4,4} \otimes \pr{g_w^{-1} \: T^j \otimes \Xi_W{}^j_{\mu\nu}
    + g'{}^{-1} \: T^6 \otimes \Xi_B{}_{\mu\nu}\psd}}
    \nonumber \\ &\hspace{0.5cm}
    + \frac{2}{27} \, \one_8 \otimes \pr{g_s^2 \: \one_3 \otimes \Ei_{5,5} + g_w^2 \: \Ei_{4,4} \otimes \iota_1 + g'{}^2 \: \Ei_{4,4} \otimes \iota_2} \otimes \one
\displaybreak[0] \\[10pt]
\label{SM-M3}
    M_3 &= \frac{1}{4} \, \Gamma^\mu \otimes \one_4 \otimes \one_5 \otimes P_\mu\psd
    - g_s \: \Gamma^\mu \otimes \lambda^u \otimes \iota_1 \otimes G^u_\mu
    - g_w \: \Gamma^\mu \Gamma_L \otimes \one_4 \otimes T^j \otimes W^j_\mu
    \nonumber \\ &\hspace{0.5cm}
    - g' \: \Gamma^\mu \Gamma_R \otimes \one_4 \otimes T^6 \otimes B_\mu\psd
    - \frac{1}{6} \, g' \: \Gamma^\mu \otimes \pr{\one_3 - 3 \: \Ei_{4,\,4}} \otimes \iota_1 \otimes B_\mu\psd
    \nonumber \displaybreak[0] \\ &\hspace{0.5cm}
    + \Ei_{\alpha,\,8+g} \otimes \Ei_{r,\,4} \otimes \pr{\Ei_{a,\,5} + \Ei_{a+2,\,5}} \otimes Q_{\alpha g}^{ra}
    \nonumber \\ &\hspace{0.5cm}
    + \sqrt{2} \pr{\Ei_{8+g,\,\alpha+4} \, \Gamma_L \otimes \Ei_{r,\,4} \otimes \Ei_{a,\,5} + \Ei_{8+g,\,\alpha+4} \, \Gamma_R \otimes \Ei_{r,\,4} \otimes \Ei_{a+2,\,5}} \otimes Q_{\alpha g}^{ra}
    \nonumber \\ &\hspace{0.5cm}
    + \Ei_{8+g,\,\alpha} \otimes \Ei_{4,\,r} \otimes \pr{\Ei_{5,\,a} + \Ei_{5,\,a+2}} \otimes \bar{Q}_{\alpha g}^{ra}
    \nonumber \\ &\hspace{0.5cm}
    + \sqrt{2} \pr{\Gamma_R \, \Ei_{\alpha+4,\,8+g} \otimes \Ei_{4,\,r} \otimes \Ei_{5,\,a} + \Gamma_L \, \Ei_{\alpha+4,\,8+g} \otimes \Ei_{4,\,r} \otimes \Ei_{5,\,a+2}} \otimes \bar{Q}_{\alpha g}^{ra}
    \nonumber \displaybreak[0] \\ &\hspace{0.5cm}
    + \Ei_{\alpha,\,8+g} \otimes \Ei_{4,\,4} \otimes \pr{\Ei_{a,\,5} + \Ei_{a+2,\,5}} \otimes L_{\alpha g}^{a}
    \nonumber \\ &\hspace{0.5cm}
    + \sqrt{2} \pr{\Ei_{8+g,\,\alpha+4} \, \Gamma_L \otimes \Ei_{4,\,4} \otimes \Ei_{a,\,5} + \Ei_{8+g,\,\alpha+4} \, \Gamma_R \otimes \Ei_{4,\,4} \otimes \Ei_{a+2,\,5}} \otimes L_{\alpha g}^{a}
    \nonumber \\ &\hspace{0.5cm}
    + \Ei_{8+g,\,\alpha} \otimes \Ei_{4,\,4} \otimes \pr{\Ei_{5,\,a} + \Ei_{5,\,a+2}} \otimes \bar{L}_{\alpha g}^{a}
    \nonumber \\ &\hspace{0.5cm}
    + \sqrt{2} \pr{\Gamma_R \, \Ei_{\alpha+4,\,8+g} \otimes \Ei_{4,\,4} \otimes \Ei_{5,\,a} + \Gamma_L \, \Ei_{\alpha+4,\,8+g} \otimes \Ei_{4,\,4} \otimes \Ei_{5,\,a+2}} \otimes \bar{L}_{\alpha g}^{a}
    \nonumber \displaybreak[0] \\ &\hspace{0.5cm}
    + \frac{1}{2} \pr{\bar{Y}^{(d)} \otimes \one_3 + \bar{Y}^{(e)} \otimes \Ei_{4,\,4}} \otimes \Ei_{a,\,4} \otimes \Phi^a
    \nonumber \\ &\hspace{0.5cm}
    + \frac{1}{2} \pr{\bar{Y}^{(d)}{}^\dagger \otimes \one_3 + \bar{Y}^{(e)}{}^\dagger \otimes \Ei_{4,\,4}} \otimes \Ei_{a,\,4} \otimes \Phi^a{}^\dagger
    \nonumber \\ &\hspace{0.5cm}
    + \frac{1}{2} \pr{\bar{Y}^{(u)} \otimes \one_3 + \bar{Y}^{(\nu)} \otimes \Ei_{4,\,4}} \otimes \Ei_{a,\,3} \otimes \tilde{\Phi}^a
    \nonumber \\ &\hspace{0.5cm}
    + \frac{1}{2} \pr{\bar{Y}^{(u)}{}^\dagger \otimes \one_3 + \bar{Y}^{(\nu)}{}^\dagger \otimes \Ei_{4,\,4}} \otimes \Ei_{a,\,3} \otimes \tilde{\Phi}^a{}^\dagger
\displaybreak[0] \\[10pt]
\label{SM-M4}
    M_4 &= \frac{1}{2} \, \Gamma^\mu \otimes \Ei_{4,4} \otimes \pr{\pr{\iota_1 - \iota_2} \otimes P_\mu\psd + \frac{2}{3} \, g_w \: T^j \otimes W_\mu^j + \frac{2}{3} \, g' \: T^6 \otimes B_\mu\psd}
    \nonumber \\ &\hspace{0.5cm}
    + \frac{1}{4} \, \Gamma^\mu \otimes \Ei_{4,4} \otimes \pr{\Ei_{a,3} \otimes \Pi^a_\mu + \Ei_{3,a} \otimes \Pi^a_\mu{}^\dagger}
    \nonumber \\ &\hspace{0.5cm}
    + \frac{1}{2} \, \one_8 \otimes \Ei_{4,4} \otimes \pr{\Ei_{a,3} \otimes \Phi^a + \Ei_{3,a} \otimes \Phi^a{}^\dagger}
\displaybreak[0] \\[10pt]
\label{SM-M5}
    M_5 &= \frac{1}{4} \, \Gamma^\mu \otimes \one_3 \otimes \pr{\Ei_{a,3} \otimes \Pi^a_\mu + \Ei_{3,a} \otimes \Pi^a_\mu{}^\dagger}
    - \frac{2}{9} \, \one_8 \otimes \one_3 \otimes \iota_1 \otimes \one
    \nonumber \\ &\hspace{0.5cm}
    + \frac{1}{2} \, \one_8 \otimes \Ei_{4,4} \otimes \bigg(
    \Ei_{a,3} \otimes \Phi^a + \Ei_{3,a} \otimes \Phi^a{}^\dagger
    + \frac{2\nu}{3} \pr{- \iota_2 + \iota_3} \otimes \xi
    \nonumber \\ &\hspace{4cm}
    + \frac{\mu^2}{3} \, \iota_1 \otimes \one
    + \frac{3}{16 \nu} \pr{- \iota_1 + \iota_2 + \iota_3} \otimes \one
    \bigg)
\end{align}
With this freezing of $M$, the cubic matrix action $S = \NTr(M^3)$ becomes equivalent to the Standard Model action on a 4-torus.

\section{Beyond the Standard Model}
\label{sec:BSM}

The reconstruction of the Standard Model as a cubic matrix theory, as we accomplished in Section \ref{sec:SM}, provides a new perspective on the naturalness of various physical postulates beyond the Standard Model, which will be the topic of this section.

\subsection{Naturalness criteria}
\label{sec:Naturalness}

Before we discuss what might lie beyond, we should ask how natural is the Standard Model freezing itself, which can then motivate some postulates as a direct improvement for its naturalness.

As partly mentioned in Section \ref{sec:YMF} and at the beginning of Section \ref{sec:SM}, the \emph{criteria for the naturalness of a freezing} in the cubic matrix framework can be stated as follows:
\begin{enumerate}
    \item[1.] The decomposition of the matrix $M$ via a basis of locators should be determined by the symmetry content of the theory, and the locators in this decomposition should have the minimum size compatible with that content.
    
    \item[2.] The freezing should consist of as few fragments as possible.
    
    \item[3.] The field matrices that appear in multiple fragments should be paired with a consistent locator reflecting its representation,
    
    \item[4.] and their coefficients should either be identical or completely independent from each other.
\end{enumerate}
To give an example for the first two criteria, the chromodynamic and electroweak locators in \eqref{SM-Loc} are minimal with respect to the SM gauge group, admitting fields in only the scalar, (anti-)fundamental or adjoint representations. With these locators and only a single fragment $M_3$ \eqref{SM-M3} to harbor the fermionic degrees of freedom, there is not much room to add arbitrary new fermions to the Standard Model (in the sense of unfreezing the corresponding entries of $M_3$). A possible exception here would be a set of $SU(2)_L$ singlets that can be added to the electroweak corner $\Ei_{5,5}$, which can be either $SU(3)_C$ triplets or singlets (see Section \ref{sec:Dark}).

To motivate the third criterion, we observe the following conceptual challenge in this framework. A fragmented matrix action, $S = S_1 + S_2 + \cdots$, can be interpreted as being fundamentally reducible into a combination of multiple actions. In that case, how can we identify two sets of variables in two different fragments as corresponding to the same physical degrees of freedom? For example, why should the gluon entries $G^u_\mu$ in the Yang-Mills fragment $M_1$ \eqref{SM-M1} be identified with the gluon entries $G^u_\mu$ in the Dirac fragment $M_3$ \eqref{SM-M3} as the same field, if they enter the SM action $S = S_1 + S_2 + S_3 + S_4 + S_5$ separately?

To address this problem, we expect that there has to be a one-to-one correspondence between the locators and the hypothetical degrees of freedom in the decomposition \eqref{SM-Loc}, which should be uniform and consistent across all fragments. Then, we can consistently postulate, for example, that the term $\Gamma^\mu \otimes \lambda^u \otimes \one_5 \otimes X^u_\mu$ in \eqref{SM-Loc} is to be identified as gluons across all fragments, whether it is frozen or unfrozen.

The fourth criterion is about the fine-tuning of coefficients. A general challenge for finding a freezing that creates a targeted action is to avoid undesired cross-terms between the variables inside a fragment (see Section \ref{sec:YMF}). This often requires fixing some coefficients in a precise way across multiple fragments, such that the undesired cross-terms cancel out. Whenever we rely on such cancellations to build an action in the cubic matrix framework, we introduce a new fine-tuning problem to the theory that does not exist in the conventional field theory formulation.

When we create a new fine-tuning problem for an existing theory in this way through our matrix reformulation, we can turn the logic around and ask about whether the cancelled terms could in fact be considered as viable additions to the theory with phenomenological prospects. We will discuss an example for such questions about the Standard Model in the next subsection.

All in all, our Standard Model construction (\ref{SM-M1}-\ref{SM-M5}) could be ranked as somewhat mediocre with respect to these naturalness criteria. This is a great situation for looking what might be beyond the SM, since this perspective reveals several concrete ways for its improvement that are worth considering.

\subsection{Cubic interaction of the gauge 2-form}
\label{sec:BSM-Urbantke}

As we mentioned in Section \ref{sec:YMF}, the Yang-Mills matrix $M_1$ in \eqref{SM-M1} creates the cubic interaction terms
\begin{align}
\label{BSM-UrbantkeTerms}
    - \frac{27}{32} \, g_s^{-3} f_{(3)}^{uvw} \NTr(\Xi_G{}^u{}^\mu{}_\nu \, \Xi_G{}^v{}^\nu{}_\rho \, \Xi_G{}^w{}^\rho{}_\mu)
    - \frac{27}{32} \, g_w^{-3} f_{(2)}^{jkl} \NTr(\Xi_W{}^j{}^\mu{}_\nu \, \Xi_W{}^k{}^\nu{}_\rho \, \Xi_W{}^l{}^\rho{}_\mu)
\end{align}
for the 2-form fields of non-Abelian gauge groups. Unlike the similar terms appearing in the quantum effective action, these are classical terms at the order $\mathcal{O}(\hbar^0)$. We cancel these terms in the Standard Model by fine-tuning the coefficient of the first line in \eqref{SM-M2}.

Now that we created ourselves a new fine-tuning problem in the Standard Model, a reasonable next step would be to relax the fine-tuning, and consider the possibility that the coefficient of the first line in \eqref{SM-M2} could be arbitrary, according to the 4th naturalness criterion. This accounts to adding to the Standard Model two extra terms that would be proportional to those in \eqref{BSM-UrbantkeTerms}.

These new terms are proportional to a negative power of the coupling constants, which implies that they introduce non-perturbative interactions. They also appear to be related to an expression for the metric that was first developed by H.~Urbantke in \cite{Urbantke}. We do not know much else about these terms.

\subsection{Interactions between the auxiliary fields}

The matrices $M_1$ \eqref{SM-M1} and $M_4$ \eqref{SM-M4} in our SM construction display some level of compatibility, which implies that it could be possible to merge them to obtain a more natural freezing with fewer fragments. However, this attempt of merging the two fragments would create some new interaction terms between the auxiliary vector field $\Pi^a_\mu$ and the auxiliary 2-forms $\Xi_W\psd{}^j_{\mu\nu}$, $\Xi_B\psd{}^j_{\mu\nu}$, all of which we had originally introduced to convert the quartic Klein-Gordon and Yang-Mills actions into a cubic expression via the first-order formalism. These new interaction terms would be proportional to
\begin{align}
    g_w^{-1} \, \tau^j{}_{ab} \NTr(\Pi^{a\mu}{}^\dagger \, \Xi_W\psd{}^j_{\mu\nu} \, \Pi^{b\nu})
    \qquad \text{and} \qquad
    g'{}^{-1} \NTr(\Pi^{a\mu}{}^\dagger \, \Xi_B\psd{}_{\mu\nu}\psd \, \Pi^{a\nu})
    \;.
\end{align}
These terms are comparable to the $\NTr(\Xi \: \Xi \: \Xi)$ terms from the previous subsection, as they similarly introduce non-perturbative interactions among the fields that had entered the model as `auxiliary'. 

\subsection{Left-right symmetric extension}

We introduced the locator $T^6$ for the electroweak group $SU(2)_L \times U(1)_Y$ in \eqref{SM-T6} in a particular way that makes it appear as the remnant of a broken right-handed symmetry group $SU(2)_R$. This was an intentional choice we made here, motivated by a desire to pack together the covariant derivatives of the left- and right-handed fermions as in \eqref{SM-CovFer}. The fact that this arrangement was possible in harmony with all the other pieces in our matrix freezing is a non-trivial aspect of the Standard Model.

Inside the $5 \times 5$ matrices that we utilized as the electroweak locators, we can roughly say that the indices one and two govern left-handed transformations, while the indices three and four govern right-handed transformations. The index five is reserved for the fermionic matter. The Higgs field $\Phi^a$, $a \in \{1,2\}$, is placed at $\Ei_{a,3}$ in \eqref{SM-M4}, which makes it an $SU(2)_L$ doublet and gives it the hypercharge $+1/2$. This placement of the Higgs field into the off-diagonal $2 \times 2$ blocks of the electroweak locators was also crucial for building the Yukawa interaction from the matrix $M_3$ in \eqref{SM-M3}.

It is particularly worth noting how the value $+1/2$ for the hypercharge of the Higgs field appears naturally in this construction. We kept the same coefficient for $B_\mu$ in \eqref{SM-M4} as that in the Yang-Mills fragment \eqref{SM-M1}, which then uniquely determines the hypercharge for a field located at $\Ei_{a,3}$. This hypercharge happened to be the correct value for the Higgs field.

It would be possible to combine this cubic matrix perspective on the Standard Model with a Grand Unification approach to get an insight on the missing degrees of freedom. Our locator choices in this paper would be suited ideally for a left-right symmetric extension, such as the Pati-Salam model $SU(4) \times SU(2)_L \times SU(2)_R$. Alternatively, it is also possible to choose a different set of locators for the Standard Model such that it would make a chiral unification more straightforward. We leave this subject for a future paper.

\subsection{Dark matter?}
\label{sec:Dark}

A tempting question here is whether this matrix reformulation of the Standard Model suggests any dark matter candidates. It does suggest a handful of new degrees of freedom and interaction terms that can be considered for unfreezing within the same universality class, hence the answer depends heavily on what kind of a dark matter candidate one is expecting to find. There is one particular candidate that would be worth mentioning here briefly.

Recall firstly that adding a new field to a cubic matrix theory $S = \NTr(M^3)$ is different from adding it conventionally to a field theory. The set of all hypothetical fields embedded inside $M$ within any fixed structure is a finite list, which is determined by the choice of locators and the number of fragments.

In particular, it attests to some degree of completeness for the Standard Model that there is an associated species of fermions both on the edge $\Ei_{r,4}$ of the chromodynamic locator (the quarks), as well as on its corner $\Ei_{4,4}$ (the leptons). Hence, all fermionic parts of the chromodynamic locator of the matrix $M_3$ are activated in the Standard Model.

The same cannot be said about the electroweak locators, as there is no known fermion species located on the corner $\Ei_{5,5}$ of our matrix. This raises questions, because these locators contain a set of variables that are \emph{already} part of the matrix $M$, but they appear to be frozen as far as the known matter fields are concerned. If we activate them into degrees of freedom, they would give rise to an $SU(2)_L$ singlet fermion field with zero hypercharge. This description fits the sterility that is necessary for any reasonable dark matter candidate. In fact, there are two possible fermion species on the electroweak corner $\Ei_{5,5}$, as they can be either $SU(3)_C$ triplets or singlets.

Any interaction between these candidate fields and the known fields can be constrained in our framework using the naturalness criteria in Section \ref{sec:Naturalness}. We also leave this subject here for future research.

\section{Conclusion}
\label{sec:Conclusion}

We presented in this paper a new approach to classical and quantum field theories via a matrix model based on infinite Toeplitz matrices, which has numerous applications and future directions that are worth highlighting here.

An explicit demonstration of the equivalence between field theories and pure matrix theories, where the spacetime dependence is replaced entirely by Toeplitz matrix indices, was missing in the literature of matrix models to the best of our knowledge\footnote{The most similar approach is found in \cite{Kazakov:2000ar}.}. Most other types of matrix theories either describe a system of D-branes, or have spacetime-dependent entries, leading to a different approach than the one we have taken here. Therefore, we hope that the framework we laid out in this paper will prove significant for building the bridges between string theory, field theory, and matrix theories.

Our matrix approach can also be used as a computational framework for field theories. In contrast to the lattice field theory approach, we preserve the continuum of the spacetime and instead discretize the energy-momentum space. This can become advantageous for performing certain non-perturbative calculations, such as the calculation of glueball masses. Representing the physical system in pure matrices may also increase the efficiency of numerical methods such as Monte Carlo simulations, since a variety of tools from linear algebra become available for use.

For both our theoretical and computational goals, a key challenge is to formulate the finite truncations of these matrix models in a consistent framework while maintaining the isomorphism between the truncated matrix and field algebras. This could shed a new light on the quantization and renormalization of field theories through matrix methods. We have ongoing work in this direction and it will be the topic of a subsequent paper.

The freezing technique links the cubic action of a single matrix $M$ to a wide range of multiple-matrix theories that correspond to realistic field theories including the Standard Model. From the particle physics' perspective, this link provides new ways of asking questions about the Standard Model structure and what might lie beyond it. Turning a given classical field action into a freezing carries a series of challenges both for generating the desired terms and for avoiding the undesired ones, which are accompanied by a new set of criteria for the naturalness of this procedure. As we discussed here in several examples, approaching the Standard Model by a matrix freezing offers a well-defined framework to motivate new ideas in high-energy physics.

One can also construct the action for general relativity as a freezing of the cubic matrix model in the Plebanski formalism \cite{Smolin:2008pk, CDJM, Krasnov:2017epi}, though coupling gravity to the matter fields under the cubic ansatz is more challenging. Ideally, we hope to recover the 4D gravity as an emergent phenomenon in these matrix models.

Finally, the freezing method provides an algorithmic procedure for organizing field theories into a landscape. One opportunity for the future would be to incorporate this method in a computer program that searches for new theories in fundamental physics and evaluates them automatically for their relevance in the real world with respect to a given set of experimental inputs. We can thus speculate of a futuristic scenario where a machine might discover a mathematical model that would prove useful for humankind's understanding of the physical world.

$$$$

\subsection*{Acknowledgements}

Microsoft supported this research both by funding researchers and providing computational, logistical and other general resources. The authors thank Kevin Scott of Microsoft in particular for support of this project.

This research was supported in part by Perimeter Institute for Theoretical Physics. Research at Perimeter Institute is supported by the Government of Canada through Industry Canada and by the Province of Ontario through the Ministry of Research and Innovation. This research was also partly supported by grants from NSERC, FQXi and the John Templeton Foundation.

We thank Stephon Alexander, Andrzej Banburski, William Cunningham, Tatsuya Daniel, Michael Freedman, Edward Frenkel, Omar Malik, Chetan Nayak, Vasudev Shyam, Stefan Stanojevic, Michael Toomey for valuable feedback and interactions.

\bibliography{References.bib}
\bibliographystyle{Utphys}

\end{document}